\documentclass[twocolumn,prb,showpacs]{revtex4}
\usepackage{psfrag}
\usepackage{graphicx}
\usepackage{amsmath}
\usepackage{amssymb}

\begin{document}
\title{Electron transport in a Pt-CO-Pt nanocontact: First-principles calculations}
\author{M. Strange}
\affiliation{Center for Atomic-scale Materials Physics, Department of Physics \\
  NanoDTU, Technical University of Denmark, DK-2800 Kgs. Lyngby, Denmark}
\author{K. S. Thygesen}
\affiliation{Center for Atomic-scale Materials Physics, Department of Physics \\
  NanoDTU, Technical University of Denmark, DK-2800 Kgs. Lyngby, Denmark}
\author{K. W. Jacobsen}
\affiliation{Center for Atomic-scale Materials Physics, Department of Physics \\
  NanoDTU, Technical University of Denmark, DK-2800 Kgs. Lyngby, Denmark}

\date{\today}

\begin{abstract}
  We have performed first-principles calculations for the mechanic
  and electric properties of pure Pt nano-contacts and a Pt contact
  with a single CO molecule adsorbed. For the pure Pt contacts we see
  a clear difference between point contacts and short chains in good
  agreement with experiments. We identify a tilted bridge
  configuration for the Pt-CO-Pt contact, which is  stable and has a
  conductance close to $0.5G_0$ ($G_0=2e^2/h$), and we propose that
  this structure is responsible for an observed peak at $0.5G_0$ in
  the conductance histogram for Pt exposed to a CO gas. We explain the
  main features of the transmission function for the Pt-CO-Pt contact,
  and show that the conductance is largely determined by the local $d$-band at
  the Pt apex atoms.
 \end{abstract}
\maketitle

\begin{section}{Introduction}\label{sec:intro}
  Advances in the experimental techniques for manipulating and
  contacting individual atoms and molecules, and the vision
  of using simple organic molecules as the basic
  building blocks in electronic devices, have recently intensified the
  interest for electron transport in nano-scale
  contacts.~\cite{aviram74,eigler91,ohnishi98,joachim00}  On the theoretical side a
  number of first principles methods to describe the electrical
  properties of realistic atomic-sized junctions have been
  developed.~\cite{transiesta,ratner,smeagol,calzolari04,fujimoto03} Many 
  of these methods combine a single-particle description of the electronic
  structure extracted from density functional theory (DFT),\cite{sham65} 
  with a non-equilibrium Green's function formalism\cite{keldysh65} 
  to calculate the current.
  The various schemes mainly differ in the choices of basis sets and the
  way in which the coupling between the molecule and macroscopic leads
  is taken into account.
  
  Regardless of the details of the implementation, the LDA/GGA DFT based
  transport schemes show the same general trend: the calculated
  conductance for molecular contacts in the low-conducting regime,
  i.e. contacts with a conductance much lower than the conductance
  unit $G_0=2e^2/h$, is 2-3 orders of magnitude higher than
  corresponding experimental
  values.~\cite{xue05,heurich02,stokbro03,derosa01,ventra00} 
  A number of reasons for this
  discrepancy have been proposed, including correlation effects not
  captured by the mean-field approach~\cite{greer04,ferretti05} and differences
  between the atomic structures used in the calculations and those
  realized under experimental conditions.~\cite{stokbro03,ventra00}
  On the other hand, a most satisfactory agreement between DFT based
  calculations and experiments is found for the conductance of
  homogeneous metallic point contacts and monatomic
  wires. For these systems, both
  the size of the conductance, the number of conductance
  channels, and the so-called conductance oscillations are well
  reproduced by calculations.~\cite{scheer98,vmgarcia,kobayashi00,transiesta,fujimoto03,ono04,jelinek03,nakamura99,palacios02,lee04}
  The same good agreement between
  theory and experiment has been found for a heterogeneous molecular
  junction consisting of a single hydrogen molecule captured between
  Pt electrodes.~\cite{smit02,thygesen_h2conductance,vmgarcia05} 
  Common to all these systems, for which DFT 
  provides a description of the transport in agreement with
  experiment, is that the coupling to the metallic leads is strong and
  the conductance is high, i.e. on the order of $1G_0$. 

  There remains much to be learned about the reasons for these apparent 
  trends in DFT based transport calculations and it is therefore 
  desirable to extent the knowledge of systems, for which
  transport properties can be treated within the framework of DFT. 
  
  In this paper we present DFT calculations for pure Pt contacts 
  and a Pt-CO-Pt system, which according to experiments
  have a conductance on the order of $1G_0$ and thus belongs to the group
  of systems for which previous DFT transport calculations 
  have been successful.
  We have calculated the total energy and
  conductances of pure Pt contacts and chains as well as for Pt contacts
  with a single CO molecule adsorbed. Our main result is the
  identification of a certain "tilted bridge" configuration for the
  Pt-CO-Pt contact which is energetically stable and has a
  conductance close to $0.5G_0$ in agreement with recent experimental
  results.\cite{untiedt04} We find that the transport properties 
  of the Pt-CO-Pt junction to a large extend are determined 
  by the properties of the bare Pt electrodes. 
  For this reason we put some emphasis on
  verifying the ability of our method to reproduce key characteristics
  of the transport properties of pure Pt contacts and chains. Also
  here we find good agreement with experiments.
  
  Mechanically controlled break junction experiments performed at
  cryogenic temperature on pure Pt samples
  show, that as a Pt contact is pulled apart a structure with a
  characteristic conductance of around $1.5G_0$ is formed in the last
  stages before the contact breaks. This is inferred from 
  conductance histograms\cite{krans93,sknielsen,untiedt04} which show
  a pronounced peak at this value. In addition to the peak at
  $1.5G_0$, many histograms on Pt contain a smaller and broader peak
  at around $2.1G_0$. The two peaks are believed to correspond to
  chains and atomic point contacts, respectively. The fact that the
  peak at $1.5G_0$ is higher than the peak at $2.1G_0$ is explained by
  the suppression of point contacts by the formation of chains.
  Experimental evidence for this hypothesis comes from conductance
  histograms recorded as the broken contacts are brought back into
  contact -- the so-called return histograms. Such histograms contain
  no contributions from chains, and show a single peak at
  $2.1G_0$.~\cite{sknielsen} Our calculations
  predict that Pt point contacts have a conductance of $(2.0-2.3)G_0$,
  whereas short Pt chains have a conductance of $(1.3-2.0)G_0$ in good
  agreement with the experimental findings.
  
  When the Pt contact is exposed to a CO gas, the peaks
  characteristic for pure Pt disappears from the conductance histogram
  and are instead replaced by two peaks at $\sim0.5G_0$ and $\sim1.1G_0$. The
  physical origins of these peaks have not been identified
  experimentally. Recent measurements on magnetic as well as
  non-magnetic metal point contacts\cite{rodrigues03}, showed
  a fractional conductance of $0.5G_0$, and this was interpreted as
  the lifting of a spin-degenerate conductance channel. However, none
  of these results could be reproduced in experiments by C.~Untiedt
  and co-workers.~\cite{untiedt04} Instead, they suggested that the
  reported fractional conductances could result from CO contamination.
  Regardless of whether CO is the source of the reported fractional
  conductances, the question of the physical mechanism of the peak at
  $0.5G_0$ in the Pt-CO-Pt histogram still remains: is it a spin effect
  or does it have some other origin ?
  
  On the basis of our calculations we propose that the $0.5G_0$ peak
  in the Pt-CO-Pt histogram is due to a configuration where a single CO
  molecule provides a "tilted bridge" between two Pt apex atoms. We
  find that this structure has a conductance just below $0.5G_0$ for a
  wide range of electrode displacements leading to a plateau in the
  conductance trace. By carrying out a Wannier function analysis we
  identify the current carrying states of the molecule as the $2\pi^*$
  CO orbitals. However, the main features of the transmission function are
  not determined by these orbitals but rather by the Pt
  $d$-band at the apex atoms. The fact that the Pt-C bond is much
  stronger than the Pt-O bond allows us to simplify a resonant level model
  for asymmetric coupling, and obtain a simple
  description of the transmission function in terms of a single
  $d$-like Pt orbital, the
  energy level of the $2\pi^*$ CO orbital, and a coupling
  strength. All the parameters of the model are extracted from the
  first principles calculations.

  The paper is organized as follows. In Sec.~\ref{sec:method} 
  we outline our Wannier function based transport scheme. In Sec.~\ref{sec:ptchains} we present the
  results for simulated conductance traces of pure Pt point contacts
  and short Pt chains. In Sec.~\ref{sec:ptandco} the Pt-CO-Pt system is
  investigated by simulating a conductance trace and identifying the
  origin of the main features of the transmission function.  
  Finally, Sec.~\ref{section:summary} contains a summary.
\end{section}

\begin{section}{Method}\label{sec:method}
  In this section we briefly review the computational methods used in
  the present study.  All total energy calculations are performed
  using a plane wave implementation of density functional
  theory.~\cite{dacapo} The nuclei and core electrons are described by
  ultra-soft pseudopotentials~\cite{vanderbilt90}, and exchange and
  correlation is treated at the GGA level using the PW91
  energy-functional.~\cite{pw91} The Kohn-Sham (KS) eigenstates are expanded
  in plane waves with a kinetic energy less than $25$Ry. Optimizations
  of all the considered structures have been performed, until 
  the total residual force is below $0.05\text{eV/{\AA}}$.
  
  The electrical conductance is evaluated within the
  Landauer-B{\"u}tikker formalism. Thus the system is divided into
  three regions: a left lead ($L$), a right lead ($R$), and a central
  region ($C$). The leads are assumed to be perfect conductors such
  that all scattering takes place in $C$. The linear response
  conductance due to scattering upon $C$ is given by
  $G=G_0T(\varepsilon_F)$, where $T(\varepsilon_F)$ is the elastic
  transmission function evaluated at the Fermi energy. 
  Here we evaluate $T(\varepsilon)$ from the
  self-consistent KS Hamiltonian, which we represent in terms
  of a basis consisting of maximally localized, partly occupied
  Wannier functions (WFs).  The WFs are constructed according to a
  recently developed method~\cite{thygesen_WFprl,thygesen_chemphys,thygesen_WFprb} in
  such a way, that the KS eigenstates are exactly reproducible in
  terms of the WFs within some specified energy window. The energy
  window is typically selected to include all eigenstates up to around
  $4$eV above the Fermi level, and thus the plane wave accuracy of the
  original DFT calculation is retained in the subsequent transport
  calculation.  The use of a localized basis set allows us to
  calculate $T(\varepsilon)$ using the general non-equilibrium Green's
  function formalism.~\cite{meir92} 
  This formalism is formulated in terms of orthogonal basis functions,
  but was recently generalized to the case of non-orthogonal 
  basis functions.~\cite{thygesen_no_current}
   
  Furthermore, the formalism is generally valid 
  for a finite bias voltage, however, in this study we shall focus on
  the linear response conductance.
\end{section}
  
\begin{section}{Pt point contacts and chains}\label{sec:ptchains}
  In this section we investigate the electrical properties of pure Pt
  contacts and short Pt chains between bulk Pt electrodes. This study
  serves a dual purpose. First, it provides a theoretical
  justification for assigning the two peaks at $1.5G_0$ and $2.1G_0$
  in the conductance histograms for Pt to chains and point contacts,
  respectively. Secondly, it allows us to test the ability of the
  calculational scheme against well established experimental results
  and other computer simulations before applying it in the study of
  the Pt-CO-Pt contact.

  \begin{figure}[!h]
   \includegraphics[width=0.50\linewidth,angle=0]{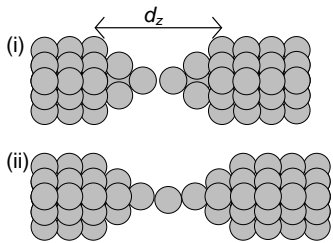}
   \caption[cap.Pt6]{\label{fig.Pt_supercell}Supercells used 
     to model the two considered structures: (i) A point-contact, and
     (ii) a 1-atom chain. The electrode
     displacement $d_z$ is defined as the distance between the (111) surfaces.
      }
    \end{figure}
    The Pt contacts are modeled using supercells by 
    two 4-atom pyramids oriented towards each other and
    attached to Pt(111) surfaces containing 3x3 atoms in the surface
    plane. Two different structures 
    are considered: (i) A point contact, 
    where the apex atoms of the pyramids are in direct contact. (ii) A 1-atom 
    chain, where a single Pt atom is inserted between the apex atoms 
    of the pyramids, see Fig. \ref{fig.Pt_supercell}. 
    In order to ensure that the effective KS potential has converged 
    to its bulk values at the end planes of the supercell we include 3-4 atomic 
    layers on either side of the pyramids.
 
    For the total energy calculations, the Brillouin zone (BZ) is
    sampled by a $4\times 4 \times 1$ Monkhort-Pack
    grid.~\cite{monkhorst} The transmission function is sampled over
    a $4 \times 4$ {\bf k}-point grid in the two dimensional BZ of the
    surface plane. This sampling is crucial in order to avoid
    unphysical features in the transmission functions due to Van Hove
    singularities associated with the quasi-one dimensional
    leads.~\cite{thygesen_kpoints}
    
\begin{subsection}{Pt point contact}
    By increasing the electrode displacement $d_z$, defined as
    the distance between the fixed (111) surfaces (see Fig.~\ref{fig.Pt_supercell}), 
    and relaxing the pyramids at each step before calculating the conductance, we simulate the process
    of creating a conductance trace.
    The result for the point contact is shown in Fig.~\ref{fig.Pt_point_trace},
    where the triangles denote conductances and the circles denote the total 
    energies measured relative to the first configuration ($d_z=10.9\text{\AA}$).
    \begin{figure}[!h]
      \includegraphics[width=0.95\linewidth,angle=0]{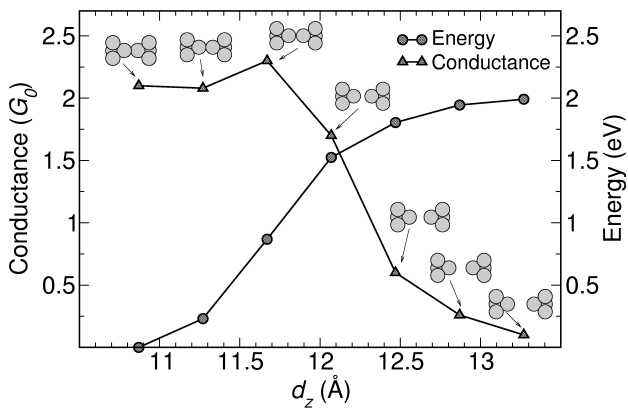}
      \caption[cap.Pt6]{\label{fig.Pt_point_trace} Conductance
        (triangles) and total energy (circles) for the Pt point contact
      as a function of electrode displacement $d_z$.
      }
    \end{figure}
    
    For the first three configurations, the conductance stays just
    above $2G_0$. The reason for the relatively weak changes in the
    conductance in this region, is that the elongation occurs quite
    uniformly over the pyramids, such that the distance between the
    apex atoms changes only slightly.  At configuration four
    ($d_z=12.07\text{\AA}$) the bond between the two apex atoms is
    broken, which can be seen directly from the insets. The breaking of the
    central bond marks the onset of a structural relaxation, which
    affects both the conductance and force significantly. Beyond this point
    the contact enters the tunneling regime and the conductance
    decreases exponentially.
    
    The simulated conductance trace is in good agreement with the experimental
    return histograms for Pt~\cite{sknielsen} which show a peak around
    $2.1G_0$. Moreover, both the
    plateau around $2G_0$ as well as the rate of the exponential decay
    in the tunneling regime compare well with the calculations
    reported in Ref.~\onlinecite{vmgarcia}.
  \end{subsection}
  \begin{subsection}{Short Pt chain}
    We have simulated the breaking of a one-atom Pt chain following the same
    procedure as for the point contacts described in the preceding
    subsection. The results are shown in Fig.~\ref{fig.Pt_1atom_trace}.
    \begin{figure}[!h]
      \includegraphics[width=0.95\linewidth,angle=0]{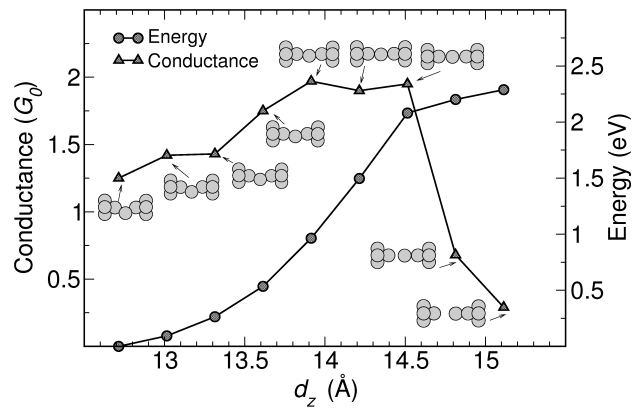}
      \caption[cap.Pt6]{\label{fig.Pt_1atom_trace} Conductance (triangles) 
        and total energy (circles) as a function of electrode 
        displacement, $d_z$. 
      }
    \end{figure}
    The calculated conductance trace has a plateau at around $1.4G_0$
    for small $d_z$ where the chain zigzags. As the contact is
    stretched further the conductance rises to a conductance just below
    $2G_0$, before
    the contact breaks at $d_z=14.5\text{\AA}$. At this point the
    structure relaxes towards the surfaces, and the conductance starts
    to decay exponentially as the tunneling regime is entered.  The
    correlation between structural relaxations and sharp changes in
    the conductance is a characteristic feature of the contact
    formation process. The effect has been observed experimentally for
    gold chains by measuring the conductances and forces
    simultaneously.~\cite{rubio96,rubio01} Furthermore, we notice that the
    increase in the conductance just before the contact breaks, seen
    in both Fig.~\ref{fig.Pt_point_trace} and
    Fig.~\ref{fig.Pt_1atom_trace}, is also observed experimentally,
    and is in fact characteristic of conductance traces
    for Pt and Al contacts.~\cite{krans93} In the case of Pt this behavior
    could be related to the linearizion of the chains which activates more
    conductance channels.~\cite{vmgarcia} A similar explanation
    has been given in the case of Al contacts.~\cite{kobayashi00,jelinek03}
    
    Finally, we stress that in order to obtain plateaus in a simulated
    conductance trace and thus predict the occurrence of peaks in a
    conductance histogram, it is necessary to allow the central atoms
    of the contact to relax in the elongation process. 
  \end{subsection}
\end{section}

\begin{section}{Pt-CO-Pt contact}\label{sec:ptandco}
  As discussed in Sec.~\ref{sec:intro}, the controlled exposure of
  the Pt contact to a CO gas changes the conductance histogram
  completely: the peaks at $1.5G_0$ and $2.1G_0$ characteristic of
  pure Pt are replaced by peaks at $0.5G_0$ and at $1.1G_0$. In order to
  understand the physical origin of the new peaks, we have carried out
  total energy and conductance calculations using the same setup as
  for the pure Pt contacts discussed in the preceding section.

  The results are summarized in Fig. \ref{fig.CO_trace},
  where we show the conductance (triangles) 
  and total energy (circles) as a function of the electrode 
  displacement $d_z$.
  \begin{figure}[!h]
    \includegraphics[width=0.95\linewidth,angle=0]{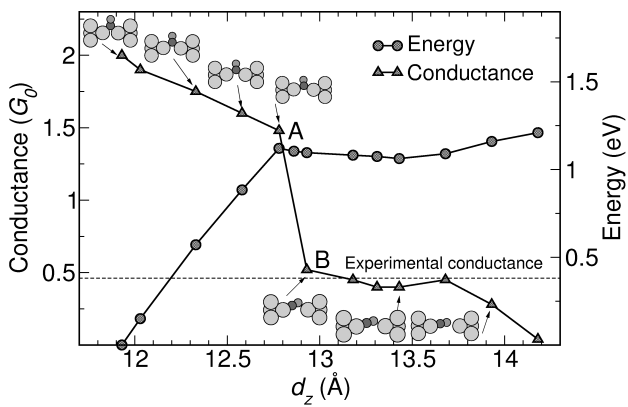}
    \caption[cap.Pt6]{\label{fig.CO_trace} Conductance (triangles) 
      and total energy (circles) 
      as a function of electrode displacement $d_z$. 
      The conductance trace is divided
      into an upper and a lower part, corresponding to the 
      CO in the upright bridge- and tilted bridge
      configuration, respectively. A and B labels the configurations just before 
      and just after CO has tilted, respectively.}
  \end{figure}
  For the initial contact geometry and small $d_z$, CO is most stable 
  at the central Pt-Pt bridge of the contact, bonding symmetrically 
  with C to the two Pt atoms in an upright bridge configuration.
  It is seen that the
  conductance decreases linearly from $2.1G_0$ to $1.5G_0$ as the
  upright bridge configuration is being stretched. At $d_z=12.8\text{\AA}$ one of
  the C-Pt bonds breaks, and the contact relaxes to a tilted bridge
  configuration. This
  qualitative change in the atomic structure is clearly seen in the
  conductance which jumps abruptly from $\sim 1.5G_0$ to a plateau at
  $\sim 0.5G_0$ which then extends over almost 1\AA.
  
  In order for the upright and tilted bridge configurations to be
  observed with a reasonable probability in the experiment, the
  binding energy of CO in these positions should be larger than that
  of CO adsorbed on the nearby sites, i.e. on the (111)-surface sites.
  This is indeed the case, as the binding energy for the upright and tilted
  bridge is around $\text{-2.6eV}$, while 
  that of CO on Pt (111) varies between $\text{-0.3eV}$ and $-1.8\text{eV}$, 
  depending on the coverage.~\cite{steckel03, doll04}
  
  Whether the conductance plateau at $0.5G_0$ in Fig.~\ref{fig.CO_trace} 
  will contribute to a peak in a conductance histogram, depends crucially 
  on the stiffness $k_s$ of the Pt electrodes.~\cite{brandbyge97} 
  Each Pt electrode 
  can be viewed as a spring connected to the CO contact, and in our setup
  only the relaxation of the atoms near to the CO molecule have been taken
  into account. These local relaxations are very important and responsible
  for the abrupt drop in the conductance curve. However, to obtain 
  a more realistic conductance
  trace, the finite spring constant of the remaining electrodes 
  should be taken into account, 
  and this will deform the $x$-axis in Fig.~\ref{fig.CO_trace}. We
  make the assumption that at some points far away from the contact,
  the position of the electrodes are controlled and denote the
  separation between these points by $L$. The relation between $L$ and
  the distance between the Pt surfaces, $d_z$, can be obtained by solving the equation
  \begin{equation}\label{eq:force_balance}
    \frac{1}{2}k_s(L-d_z) = \frac{\partial E(d_z)}{\partial d_z},
  \end{equation}
  which expresses the force balance between the springs (left hand side)
  and the contact region (right hand side). The factor $1/2$ is due to
  the fact that each springs are acting on both sides of the contact, with
  a stiffness of $k_s$.
  \begin{figure}[!h]
    \includegraphics[width=0.85\linewidth,angle=0]{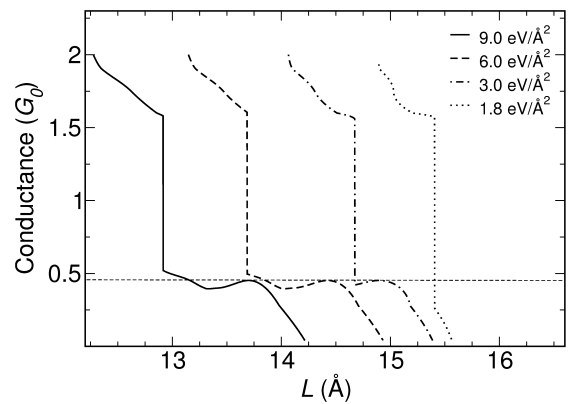}
    \caption[cap.Pt6]{\label{fig.Pt_traces_springs} Conductance traces for different
      values of the electrode stiffness, $k_s$. The size of
      $k_s$ determines the length of the conductance plateau at
      $0.5G_0$: a stiff electrode (large $k_s$) results in a long
      plateau, while a soft electrode (small $k_s$) results in a short
      or no plateau.}
  \end{figure}
  
  In Fig. \ref{fig.Pt_traces_springs}, we show the corrected
  conductance traces for $k_s$ between 1.8~eV/$\text{\AA}^2$ and 9.0~eV/$\text{\AA}^2$. 
  For $k_s$ below $\sim 2$~eV/$\text{\AA}^2$, the $0.5G_0$ conductance plateau is completely absent and
  the tilted bridge configuration would not contribute to the
  conductance histogram. However, when $k_s>3$~eV/$\text{\AA}^2$ the $0.5G_0$
  plateau becomes clearly visible, and would lead to a peak at $0.5G_0$
  in the conductance histogram. Experimental estimates for $k_s$ do
  not exist for Pt electrodes, however, in the case of gold values in the range 
  0.2eV/$\text{\AA}^2$ to 4eV/$\text{\AA}^2$ have recently been reported, 
  on the basis of non-exponential distance dependence 
  of the tunneling current.~\cite{rubio04}
   
  We therefore propose, that the observed 
  peak in the conductance histogram for CO in Pt nanocontacts is 
  due to the tilted bridge configuration. 
  All our calculations are averaged over
  spin, and the special conductance of $0.5G_0$ is 
  therefore not related to spin.
  This is further illustrated below, where we study 
  the transport mechanism in more detail. 
  
  Finally, we mention that the peak around $1.1G_0$ in the
  experimental histograms, cannot be explained by our calculations for
  a single CO molecule in the Pt contact. We have also calculated the
  conductance of the CO bridge in the one atom chain in search of a
  structure that can explain the peak at $1.1G_0$, however, we found
  conductances quite similar to those obtained for CO in the Pt point
  contact.
  
  \begin{subsection}{Conduction mechanism}
    In this section we address the question of the physical origin of
    the fractional conductance of $0.5G_0$ found for the tilted bridge
    configuration.  Fig.~\ref{fig.CO_trans_A_B} shows a typical
    transmission function for the upright bridge configuration (dashed
    line) and tilted bridge configuration (full line) immediately before and after
    the bridge tilts -- configurations A and B in
    Fig.~\ref{fig.CO_trace}. The dotted line denotes the transmission function
    obtained  for configuration B with a localized atomic-like basis set
    taken from the DFT code Siesta.~\cite{siesta} The transmission function has
    been averaged over the same k-points and calculated using the same
    electron transport code as the WFs transmission functions.
    The basis set used is the Siesta "default"
    double zeta with polarizations functions (DZP) in combination with 
    Troullier-Martins pseudopotentials.~\cite{TM} Exchange and correlation
    are described by the PBE functional,\cite{PBE} which is close to PW91 used in the
    WFs calculations.
    The transmission functions obtained using the two 
    different DFT-codes are seen to be in good
    agreement, especially in the important region near the Fermi-level.
    We have observed similar good agreements for the other Pt-CO-Pt
    structures considered in this work. This indicates that our results are independent
    of the basis set and other technical details related to the calculation
    of the self-consistent KS-Hamiltonian.
    
    \begin{figure}[!h]
      \includegraphics[width=0.90\linewidth,angle=0]{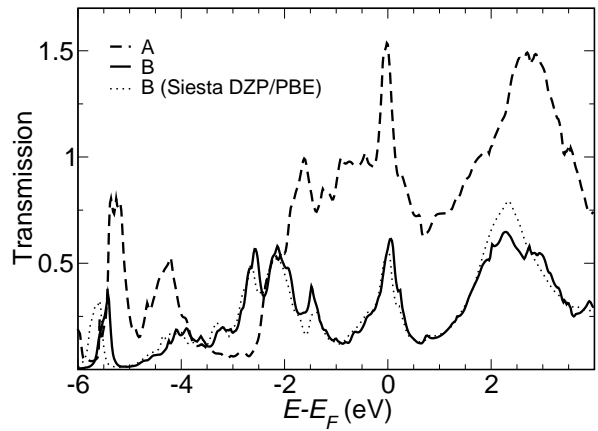}
      \caption[cap.Pt6]{\label{fig.CO_trans_A_B} Typical transmission function
        for the upright bridge (configuration A in Fig. \ref{fig.CO_trace})
        and the tilted bridge (configuration B). The dotted line
        is the transmission function obtained using Siesta-DZP/PBE.
        Both transmission functions (A and B)
        have a resonance at the Fermi level which can be related to the 
        density of states of the so-called group orbitals.
      }
    \end{figure}
    In both cases (A and B) there is a pronounced resonance at the Fermi level,
    which clearly is responsible for the conductance of $~1.5G_0$ and
    $~0.5G_0$, respectively. In fact this resonance is present for all
    the configurations investigated in this study, and it therefore
    represents a very general and robust feature of the transport
    through a Pt-CO-Pt contact. As the following analysis will show, the
    resonance is not due to the CO molecule alone, but rather is a
    (local) band structure effect related to the $d$-orbitals at the Pt
    apex atoms.
    
    To address the origin of the resonance, we perform an analysis of
    the local electronic structure in the contact region by 
    diagonalizing the Hamiltonian within the subspace spanned by the
    WFs located at the CO molecule. The orbitals and eigenvalues
    obtained in this way represent renormalized energy levels of the CO
    molecule including the effect of the coupling to the Pt leads. For
    all the considered contact geometries we find seven WFs located at
    the CO molecule, leading to seven renormalized molecular energy
    levels. For simplicity we focus on the tilted CO bridge in the following
    (configuration B), however, the conclusions are also valid for
    the upright bridge configuration. Since CO has 10 valence
    electrons, the seven renormalized CO orbitals represent
    the five occupied and the two lowest unoccupied 
    molecular orbitals. The latter are the $2\pi^*$ orbitals 
    which are known to be important for the chemisorption properties of 
    CO.~\cite{hammer96,blyholder64,doll04} By repeating the conductance
    calculations with the renormalized
    $2\pi^*$ orbitals removed from the basis set, we find that the
    resonance at the Fermi level is completely gone and the conductance 
    is reduced to $\sim
    0.05G_0$. This allows us to focus exclusively on the $2\pi^*$
    CO orbitals when analyzing the transport properties of the Pt-CO-Pt contacts.
    In the following we will refer to the $2\pi^*$ states as $|a\rangle$
    and $|b\rangle$. The on-site energies of these renormalized orbitals
    are $\varepsilon_a=1.5\text{ eV}$ and 
    $\varepsilon_b=1.6\text{ eV}$, respectively. The splitting of the
    levels is induced by the different couplings to the electrodes.

    Each of the molecular orbitals (MO) $|a\rangle$ and $|b\rangle$ give
    rise to one transmission channel through the CO molecule. If we neglect
    tunneling due to direct coupling between the Pt apex atoms, and
    neglect interference between the two transport channels, we can
    analyze the problem by considering the transport through each MO
    separately. We do this using the well known single-level model for
    resonant transport. A particularly simple form of the single level
    model, is obtained by introducing the so-called group orbitals.~\cite{kwjsbog}
    The group-orbital for MO $|a\rangle$ (or $|b\rangle$) of lead 
    $\alpha$ ($\alpha=L,R$) is defined as
    $|g^a_{\alpha} \rangle=\frac{1}{V_{\alpha,a}} P_{\alpha} H |a\rangle$,
    where $H$ is the Hamiltonian, $P_{\alpha}$ is
    the orthogonal projection onto lead $\alpha$, and $1/V_{\alpha,a}$ a
    normalization constant. The name "group orbital"
    refers to the fact that $|g^a_{\alpha} \rangle$ consists of the
    group of states in the lead to which $|a\rangle$ is most strongly 
    coupled. In fact, the coupling between $|a\rangle$ and any state in lead
    $\alpha$ orthogonal to $|g^a_{\alpha} \rangle$ is zero, while the
    coupling to the group orbital is given by $V_{\alpha,a}=\langle
    a|H|g_{\alpha}\rangle$. In the limit of strong asymmetric coupling,
    when $V_{R,a}^2<<V_{L,a}^2$, the transmission function takes the form
    \begin{equation}\label{eq:trans}
      T(\varepsilon) = 4\pi^2V_{R,a}^2\rho^0_{R,a}(\varepsilon)\rho_a(\varepsilon),
    \end{equation}
    where $\rho_a(\varepsilon)$ is the projected density of states
    (PDOS) of the MO $|a\rangle$ and $\rho^0_{R,a}(\varepsilon)$ is the PDOS
    of the group orbital of the right lead in the absence of coupling to
    $|a\rangle$, i.e. calculated with $V_{R,a}=0$. The limit of strong
    asymmetric coupling is relevant for the tilted bridge configuration
    where $V_{R}^2/V_{L}^2\approx 0.1$ for both MOs $|a\rangle$ and
    $|b\rangle$. The large asymmetry in the coupling strengths indicates that
    the Pt-C bond is much stronger than the Pt-O bond. To illustrate the
    situation, the inset of Fig.~\ref{fig.group_dos} shows an iso-surface
    plot of $|a\rangle$ (transparent) together with its left and right group
    orbitals (solid). The latter consists mainly of $d$-like orbitals
    centered on the apex Pt atoms. The coupling strengths $V_{L,a}$ and
    $V_{R,a}$ are indicated.
    
    \begin{figure}[!h]
      \includegraphics[width=0.95\linewidth,angle=0]{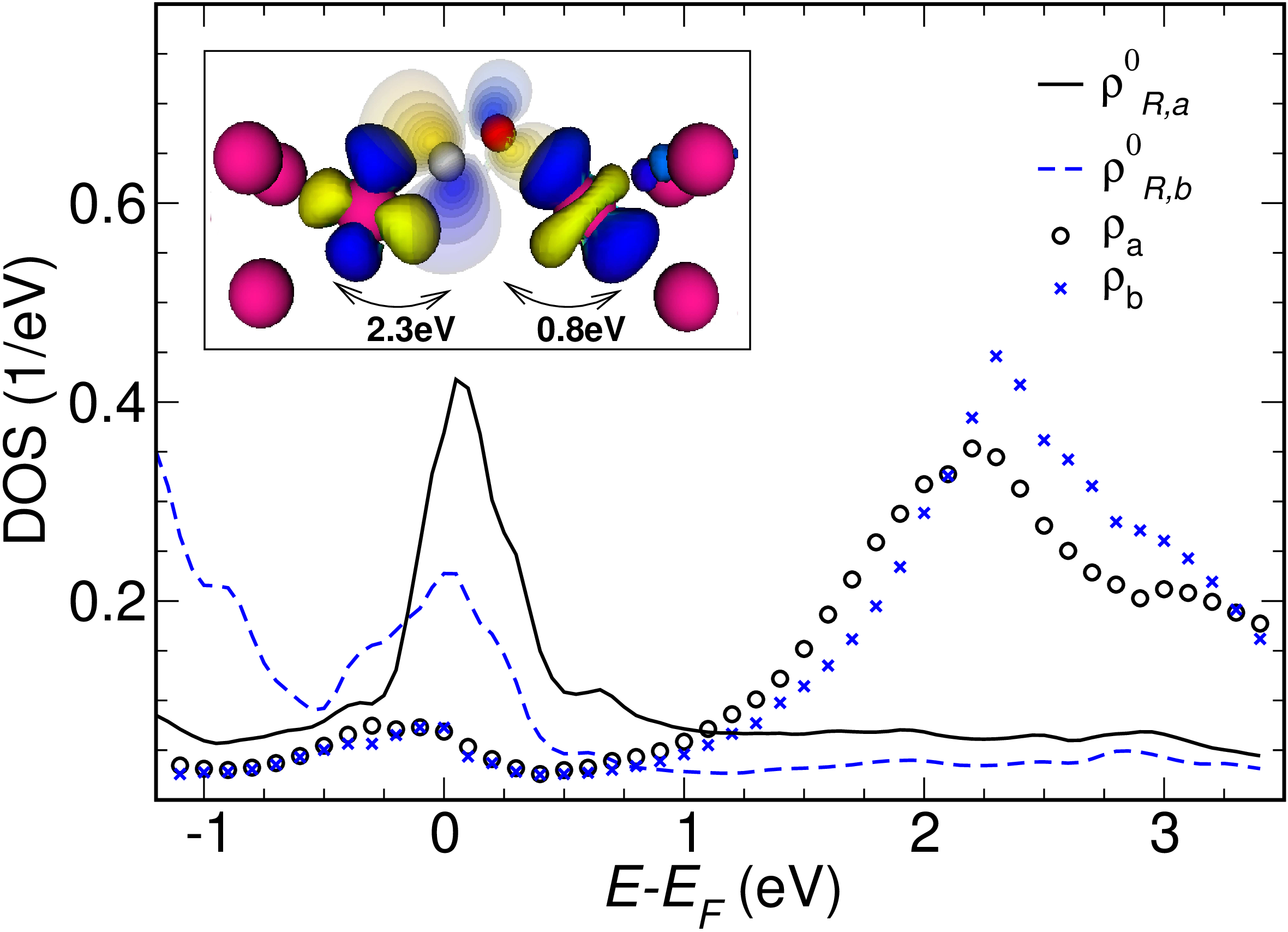}
      \caption[cap.Pt6]{\label{fig.group_dos} (Color online) The PDOS
        of the $2\pi^*$ states $|a\rangle$ and $|b\rangle$ together with
        the PDOS of the group orbital in the weakly coupled right lead.
        The inset shows an iso-surface plot of the MO $|a\rangle$
        (transparent) and its corresponding left and right group
        orbitals (solid). The PDOS for the MOs (circles and stars) is
        quite flat around the Fermi level, while the the PDOS of two
        group orbitals (full and dashed lines) both have a peak at the
        Fermi level. It is this peak that gives rise to the resonance in
        the transmission function.}
    \end{figure}  
    
    According to Eq.~(\ref{eq:trans}) the energy dependence of the
    transmission function is determined by the product of the PDOS of
    the MO and the PDOS of the group orbital in the weakly coupled lead.
    In Fig.~\ref{fig.group_dos} we show the calculated PDOS for the MOs
    $|a\rangle$ and $|b\rangle$ together with the PDOS of the
    corresponding group orbitals $|g^a_R\rangle$ and $|g^b_R\rangle$. It
    is now clear that the transmission resonance at the Fermi level
    results from a corresponding peak in the PDOS of the group orbitals
    of the right lead, or, equivalently, from a peak in the PDOS of the
    $d$-states at the Pt apex atoms. The bare energies of the MOs at 
    $\varepsilon_a=1.5$eV and $\varepsilon_b=1.6$eV, respectively, are shifted upwards by
    the coupling to the Pt $d$-band, and can be seen as broad peaks in
    the PDOS at $\sim 2.2\text{eV}$. These peaks are also clearly visible  
    in the transmission function in Fig.~\ref{fig.CO_trans_A_B}.

    The transmission resonance at $E_F$ is thus caused by the properties of 
    the isolated Pt lead, while the role of 
    the 2$\pi^*$ CO orbitals is to provide a flat background at the Fermi level and a
    peaked structure at $\sim 2.2\text{eV}$.
    Since the group orbitals for different
    configurations are very similar (always $d$-like) the determining factors in the
    transmission functions stay almost constant, and this explains the
    robustness of the main features in the transmission function.
    
    It is well known, that the calculated HOMO-LUMO gap of CO is 
    somewhat sensitive to the applied exchange-correlation functional.~\cite{doll04}
    However, since the transport properties of the investigated Pt-CO-Pt
    bridge involves only the tails 
    of the PDOS of the CO orbitals and is dominated by the PDOS of the
    Pt leads, the exact positions of the CO energy levels are not expected to be crucial.
    An accurate description of Pt is, however, important and as discussed in 
    Sec.~\ref{sec:ptchains} our results for Pt are in good agreement with
    experiments.
  \end{subsection}
\end{section}

\begin{section}{Summary}\label{section:summary}
  We have performed calculations within the framework
  of density functional theory for the mechanical and
  electrical properties of pure Pt nano-contacts and
  Pt contacts with a single CO molecule.
  
  For the pure Pt contacts, we obtain conductances traces
  which are in good agreement with experiments as well
  as other recent theoretical calculations. Our results show that Pt
  point contacts have a conductance in the range $(2.0-2.3)G_0$ while
  that of short Pt chains is $(1.3-2.0)G_0$. This provides a
  theoretical justification for assigning the peaks at $\sim 2.1G_0$
  and $\sim 1.5G_0$ in the conductance histogram for Pt to point
  contacts and chains, respectively.

  For the Pt-CO-Pt contact we identify an energetically stable configuration with the CO
  molecule providing a tilted bridge between two Pt apex atoms. Based on
  realistic DFT simulations of the creation of a conductance
  trace with the elastic response of the electrodes included through
  effective spring constants, we propose that the tilted CO bridge is
  responsible for the peak at $0.5G_0$ observed recently in the conductance
  histogram for Pt-CO-Pt. We characterize and explain the main features
  of the transmission function for the Pt-CO-Pt contact in terms of
  the properties of the isolated CO molecule and the free Pt leads. The
  analysis shows that the conductance to a large extent is determined
  by the local $d$-band at the Pt apex atoms and to a smaller extent
  by the $2\pi^*$ CO orbitals. 
\end{section}

\begin{section}{Acknowledgments}
  We would like to thank J. van Ruitenbeek and D. Djukic for illuminating 
  discussions on the interpretation of their break-junction experiments.
  We acknowledge support
  from the Danish Center for Scientific Computing through Grant No.
  HDW-1101-05.
\end{section}


\bibliographystyle{apsrev}

\end{document}